\documentclass{sigchi}

\CopyrightYear{2020}
\makeatletter
\def\@copyrightspace{\relax}
\makeatother

\usepackage{balance}       % to better equalize the last page
\usepackage{graphics}      % for EPS, load graphicx instead 
\usepackage[T1]{fontenc}   % for umlauts and other diaeresis
\usepackage{txfonts}
\usepackage{mathptmx}
\usepackage[pdflang={en-US},pdftex]{hyperref}
\usepackage{color}
\usepackage{booktabs}
\usepackage{textcomp}

\usepackage{microtype}        % Improved Tracking and Kerning
\usepackage{ccicons}          % Cite your images correctly!
\usepackage{todonotes}

\usepackage{array}
\newcolumntype{L}[1]{>{\raggedright\let\newline\\\arraybackslash\hspace{0pt}}m{#1}}
\newcommand{\squeezeup}{\vspace{-0.5cm}}
\newcommand{\squeezeupsmall}{\vspace{-0.2cm}}
\usepackage{balance}
\usepackage{cite}
\def\plaintitle{A Game-Based Approach for Helping Designers Learn Machine Learning Concepts}

\def\emptyauthor{}
\def\plainkeywords{Interactive Visualization, Game Metaphors, eXplainable AI, Human-Centered Design}

\makeatletter
\def\url@leostyle{%
  \@ifundefined{selectfont}{
    \def\UrlFont{\sf}
  }{
    \def\UrlFont{\small\bf\ttfamily}
  }}
\makeatother
\def\pprw{8.5in}
\def\pprh{11in}

\definecolor{linkColor}{RGB}{6,125,233}
\hypersetup{%
  pdftitle={\plaintitle},
  pdfauthor={\emptyauthor},
  pdfkeywords={\plainkeywords},
  pdfdisplaydoctitle=true, % For Accessibility
  bookmarksnumbered,
  pdfstartview={FitH},
  colorlinks,
  citecolor=black,
  filecolor=black,
  linkcolor=black,
  urlcolor=linkColor,
  breaklinks=true,
  hypertexnames=false
}

\begin{document}

\title{\plaintitle}

\numberofauthors{3}
\author{%
  \alignauthor{%
    \textbf{Chelsea M. Myers}\\
    \affaddr{Drexel University} \\
    \affaddr{Philadelphia, PA, USA} \\
    \email{chel.myers@gmail.com} }\alignauthor{%
    \textbf{Jiachi Xie}\\
    \affaddr{Drexel University} \\
    \affaddr{Philadelphia, PA, USA} \\
    \email{jx87@drexel.edu} } \vfil \alignauthor{%
    \textbf{Jichen Zhu}\\
    \affaddr{Drexel University} \\
    \affaddr{Philadelphia, PA, USA} \\
    \email{jichen.zhu@gmail.com} } }

\maketitle

\begin{abstract}

Machine Learning (ML) is becoming more prevalent in the systems we use daily. Yet designers of these systems are under-equipped to design with these technologies. Recently, interactive visualizations have been used to present ML concepts to non-experts. However, little research exists evaluating how designers build an understanding of ML in these environments or how to instead design interfaces that guide their learning. In a user study (n=21), we observe how designers interact with our interactive visualizer, \textit{QUBE}, focusing on visualizing Q-Learning through a game metaphor. We analyze how designers approach interactive visualizations and game metaphors to form an understanding of ML concepts and the challenges they face along the way. We found the interactive visualization significantly improved participants' high-level understanding of ML concepts. However, it did not support their ability to design with these concepts. We present themes on the challenges our participants faced when learning an ML concept and their self-guided learning behaviors. Our findings suggest design recommendations for supporting an understanding of ML concepts through guided learning interfaces and game metaphors.
\end{abstract}

% ACM Classfication


% Author Keywords

% Print the classficiation codes
% Please use the 2012 Classifiers and see this link to embed them in the text: \url{https://dl.acm.org/ccs/ccs_flat.cfm}

\section{Introduction}

% 1. What is the broader issue that your project will address [CONTEXT/MOTIVATION]
Machine Learning is increasingly recognized as a new design material for interactive application ~\cite{Dove2017, Amershi2014, Yang2018ux}. For instance, e-commerce systems use it to personalize products that match consumers' needs~\cite{ricci2011introduction}. Modern virtual assistants use it to process users' voice  dialogue. In computer games, ML is used to generate new levels in games such as {\em Super Mario Brothers}~\cite{snodgrass2016learning} and to automatically test games~\cite{holmgard2018automated}. 

Despite the growing need for ML-based user experience, designers are generally ill-equipped to understand how ML functions and to use it in their designs.

By supporting designers in understanding ML concepts, we equip them in being able to ``purposefully use ML to solve the right, user-centered problems.''~\cite{Yang2017} 

However, equipping designers is not an easy task as even expert programmers are found to struggle with ML concepts as algorithms become more sophisticated~\cite{Patel2008}. In the last decade, there has been a large increase in research employing a human-centered approach to aiding experts, such as ML researchers and programmers, in understanding ML and AI concepts~\cite{Amershi2014, Carter2019, Cai2019, Kim2017, Adadi2018, Abdul2018}. Relatively less work has been done to aid non-experts, especially designers~\cite{Zhu2018}. Among existing approaches to educate non-experts, a prominent one is to use web-based interactive visualization tools to simulate the operation of a particular ML algorithm or a family of algorithms. Popular ones include {\em Tensorflow Playground}~\cite{smilkov2017direct}, {\em Teachable Machine}~\cite{Carney2020TeachableClassification}, and {\em GANLab}~\cite{Kahng2019}. Despite the popularity of these sandbox interactive visualizers, unguided learning has been found to under-perform compared to guided learning techniques (e.g., worked examples) when teaching a variety of concepts~\cite{Kirschner2006WhyTeaching}. However, when it comes to designers, we know of little research exploring how to design and structured guide learning interfaces for ML concepts.

% We currently know little about how structure learning for designers to improve understanding of ML concepts. designers interact with them and how effective they are in supporting them in understanding ML concepts. 

% Furthermore, existing systems explore ML concepts in the context of algorithms, instead of design tasks. %{\color{red}{This creates a mental gap between [...] the context of the interactive visualization and how designers can apply the concepts presented to their own work.}}

In this paper, we present our progress in a human-centered design research project to develop an interface to educate ML concepts to designers. Specifically, we report our findings on how designers interact with an interactive visualizer designed to familiarize designers with the Q-Learning Reinforcement Learning (RL) algorithm~\cite{watkins1992q}. We designed an interactive visualization based on existing popular platforms to observe the challenges designers face when understanding our RL concept in an unguided environment. Our goal was to review these challenges in order to design more informed techniques for future guided learning interfaces. We selected RL since it is a widely used type of ML technique~\cite{Li2019ReinforcementApplications}, and Q-Learning is a classic RL algorithm. For the purpose of this study, we conduct our evaluation by observing designers interact with {\em QUBE}\footnote{available at: \url{http://jiachixie.com/Qube/qube.html}}. \textit{QUBE} adopts some of the key design features in existing interactive visualizations shared in the platforms mentioned above, namely 1) unguided exploratory learning in a sandbox environment and 2) simulation-based interactive visualization with changeable parameters. Unlike the previously mentioned interfaces, {\em QUBE} uses game metaphors to teach Q-Learning in the context of designing a 2D game level (i.e., a maze) for an agent to solve. We apply the common game metaphor of the user manipulating an ``agent'' to get through a game ``level''~\cite{Kazimoglu2012LearningGame-play,Shim2017TheStudents,Weintrop2012RoboBuilder:Game, Chao2016ExploringEnvironment}.

In our user study ($n=21$) with mixed methods, we aimed to understand how digital designers approach the interactive visualization and game metaphors of {\em QUBE}. We also analyze what gaps in their understanding can potentially be addressed in guided learning interfaces. Our results show that participating designers' self-efficacy, ML understanding, and Q-Learning understanding increased significantly after using {\em QUBE}. However, while participants had a high-level understanding of the concepts presented, the interactive visualization did not support their ability to design with these concepts effectively. Participants were unsure how to alter parameters of Q-Learning to achieve their design goals. For our game metaphor, we also observed a strong impact of the metaphors used in {\em QUBE} on participants' understanding of RL and their self-defined goals. For example, participants overestimated the Q-Learning agent's sophistication and assumed there were hidden characteristics not revealed in the interface. These challenges limited the participants' exploration of other agent and level designs. 
% Based on our findings, we present design recommendations for addressing these obstacles and develop guided learning systems to educate ML concepts to designers.

% This paper's discussion presents our design guidelines for designing guided learning experiences or game metaphors to educate ML concepts to designers. 

% We observe how this agent game metaphor impacts designers' education when the agent is slightly autonomous and powered through RL. 
% designers' understanding of ML concepts can be limited by the domain they are presented in and influenced by the implicit goals of said domain. Additionally, we present how designers structure their exploration of \textit{QUBE} and the features that inhibit their experimentation. Finally, we found designers struggled to grasp the training process necessary for Q-Learning and their role in it.

% struggle to reach their self-defined goals with the IV due to their lack of ability to manipulate parameters to achieve a desire effect. 
%  the lack of visibility of a parameter's effect and how parameters effect each other inhibited our participants ability to reach their self-defined goals with the system.

Existing work reports non-experts struggle to achieve low-level technical understandings of ML concepts in general~\cite{Dove2017, Yang2018, Yang2018InvestigatingLearning}. While confirming these findings, we further this line of research by analyzing what interface elements hinder designers from improving their technical understandings. The core contribution of this paper is as follows. To the best of our knowledge, our work is among the first empirical study of how designers interact with an interface designed specifically for their ML education. Analyzing data from designers' interactions, we derived major themes on the successes and shortcomings of interactive visualization for RL education. These themes covers aspects of general design features shared by this genre of simulation-based interactive visualization systems for ML and the application of game metaphors. This paper presents design recommendations for future guiding learning interfaces of ML, such as limiting the overestimation of a concept's sophistication, highlighting the impact of combining of parameters, and educating parameters in the context of design goals. Compared to the prior work researchers have completed with previous versions of \textit{QUBE}~\cite{Xie2019}), this paper conducts a formal evaluation with designers. For the rest of the paper, we first discuss relevant research. Next, we present \textit{QUBE} and our user study design. Finally, we present our findings and discuss their implications.

\begin{figure}
    \centering
    \includegraphics[width=8.5cm]{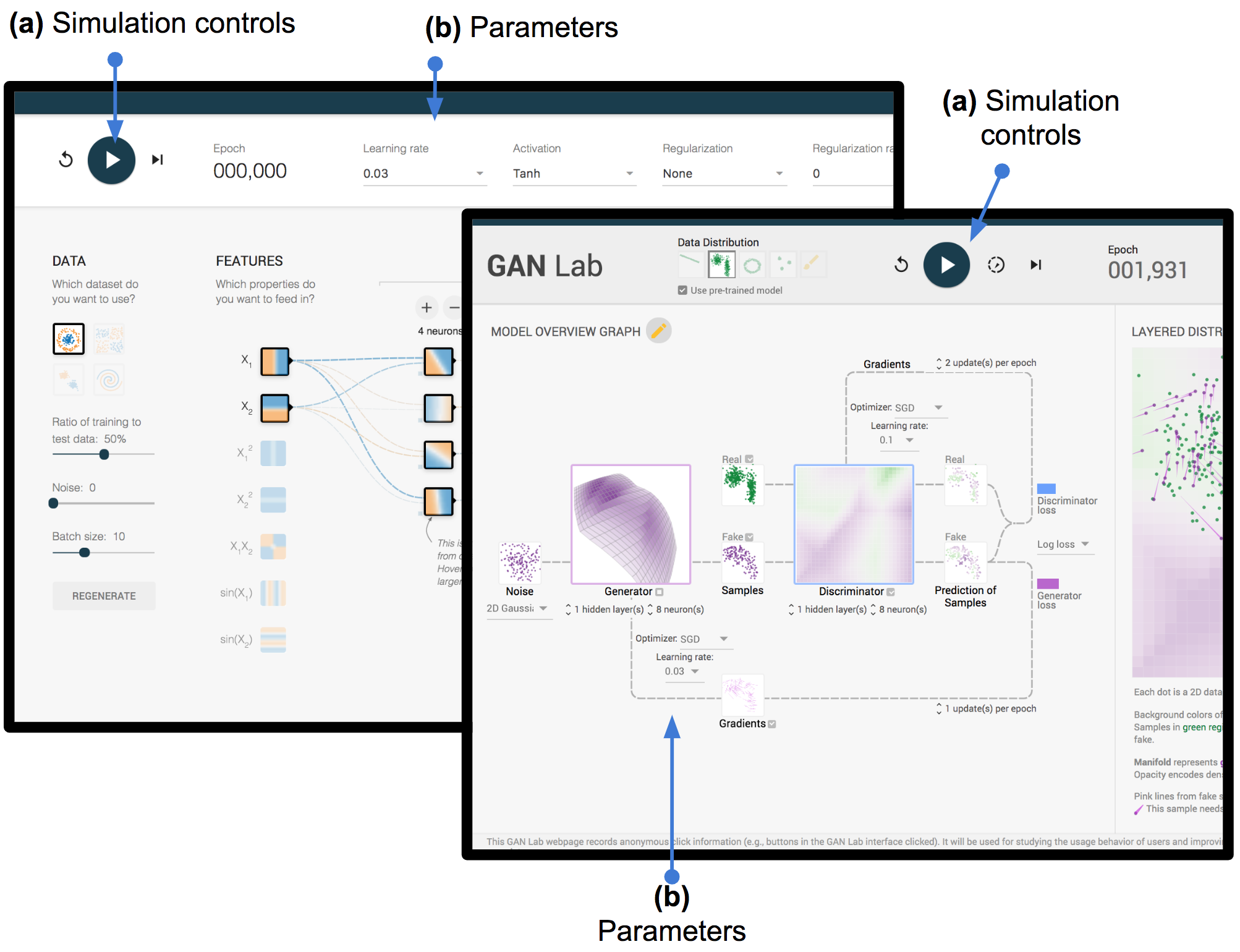}
    \caption{(a) controls for the interactive simulation and (b) changeable parameters for \textit{Tensorflow Playground} (left) and \textit{GANLab} (right)}
    \label{fig:tensorflow}
    \squeezeup
\end{figure}

%--------------------------------------------
% RELATED WORK
%--------------------------------------------
\section{Related Work}
% - a paragraph on how is ML becoming a design material for UI/UX/Game designers
ML can be leveraged by UI/UX designers to power applications more common in our current lives, such as personalized interfaces~\cite{Moon2015AdaptiveInterface, Gulla2017AdaptiveApproach, Samara2008} and conversational agents~\cite{Ganu2019AgentBuddy:Agents}. To employ ML tactics in ways currently not seen in commercial design, designers of these systems need support in understanding ML so they can use ML as a design material~\cite{Yang2018MappingInnovation,Yang2018ux, Dove2017, Zhu2018}. For a majority of existing ML tools, however, foundation knowledge of ML is required to understand the underlying processes to manipulate them. 

As ML becomes more sophisticated, increasing demands are made to make their operation more transparent and explainable to human users. The growing area of research of eXplainable AI (XAI)~\cite{Adadi2018, Abdul2018} can inform designing interactive visualizations for ML. By generating good explanations, XAI can increase the transparency of a system and the users' trust in it, their ability to interact with it, and their ability plan with it~\cite{Fox2017ExplainablePlanning}. Existing XAI techniques for non-experts focus on explaining text classifications~\cite{Stumpf2009, martens2013explaining, Lee:2017:AMG:2998181.2998230, Ribeiro2016}, translations~\cite{green2014predictive}, and context-aware systems~\cite{bellotti2001intelligibility, lim2012improving}. All of these explanations could be used in interactive visualizations to make ML concepts more transparent. With few exceptions \cite{Zhu2018b}, little XAI work focuses on designing explanations for digital designers. Designers are a special non-expert population because 1) they often have computational literacy, and 2) they need a conceptual understanding of ML to be able to effectively to incorporate it into their designs~\cite{Dove2017}. Furthermore, existing XAI research focuses more on creating new algorithms to increase AI's explainability, rather than understanding the usability and interpretability by targeted users~\cite{Abdul2018,Doshi-velez2017TowardsLearning, Miller2017ExplainableSciences}. HCI researchers are calling for a human-centered approach empowering non-experts to design with ML~\cite{Amershi2014, Xu2019}. More recently, Carney et al. released \textit{Teachable Machine}\cite{Carney2020TeachableClassification}, an interactive visualization allowing non-experts to train classification models. While receiving positive reviews by educators, it is unknown what interface elements specifically help or harm non-experts in forming their ML understanding. This paper contributes one of the few empirical evaluations of an ML interactive visualization designed through a human-centered approach which targets non-experts. Additionally, because of the abundance of XAI research, our design recommendations strive to help future researchers navigate the large field of XAI to select techniques for more guided learning interfaces. 

Prominent relevant work evaluates how UX designers understand ML concepts through \textit{interview studies}. Results have shown that UX designers find it challenging to understand ML capabilities, how to use it as a design material, and how to implement it purposefully~\cite{Dove2017}. It is also speculated designers would benefit from improved data literacy, abstracted presentations of ML, and sensitizing concepts~\cite{Yang2018InvestigatingLearning}. Zhu et al.~\cite{Zhu2018} propose a new area of research, {\em eXplainable AI for Designers,} which focuses on creating explainable AI techniques specifically for \textit{game} designers. Yang et al. interviewed non-experts who currently used ML tools and found they were susceptible to several pitfalls when building accurate ML models~\cite{Yang2018}. To our knowledge, relatively little work exists on how designers \textit{actually} interact with platforms designed to \textit{explain} ML concepts and the quality of the understanding they build as a result of said interaction. Results from this paper complement the findings from the above-mentioned interview studies and attempt to bridge this knowledge gap.

\begin{figure*}[h]
    \centering 
       \squeezeup
          \squeezeup
    \includegraphics[width=16cm]{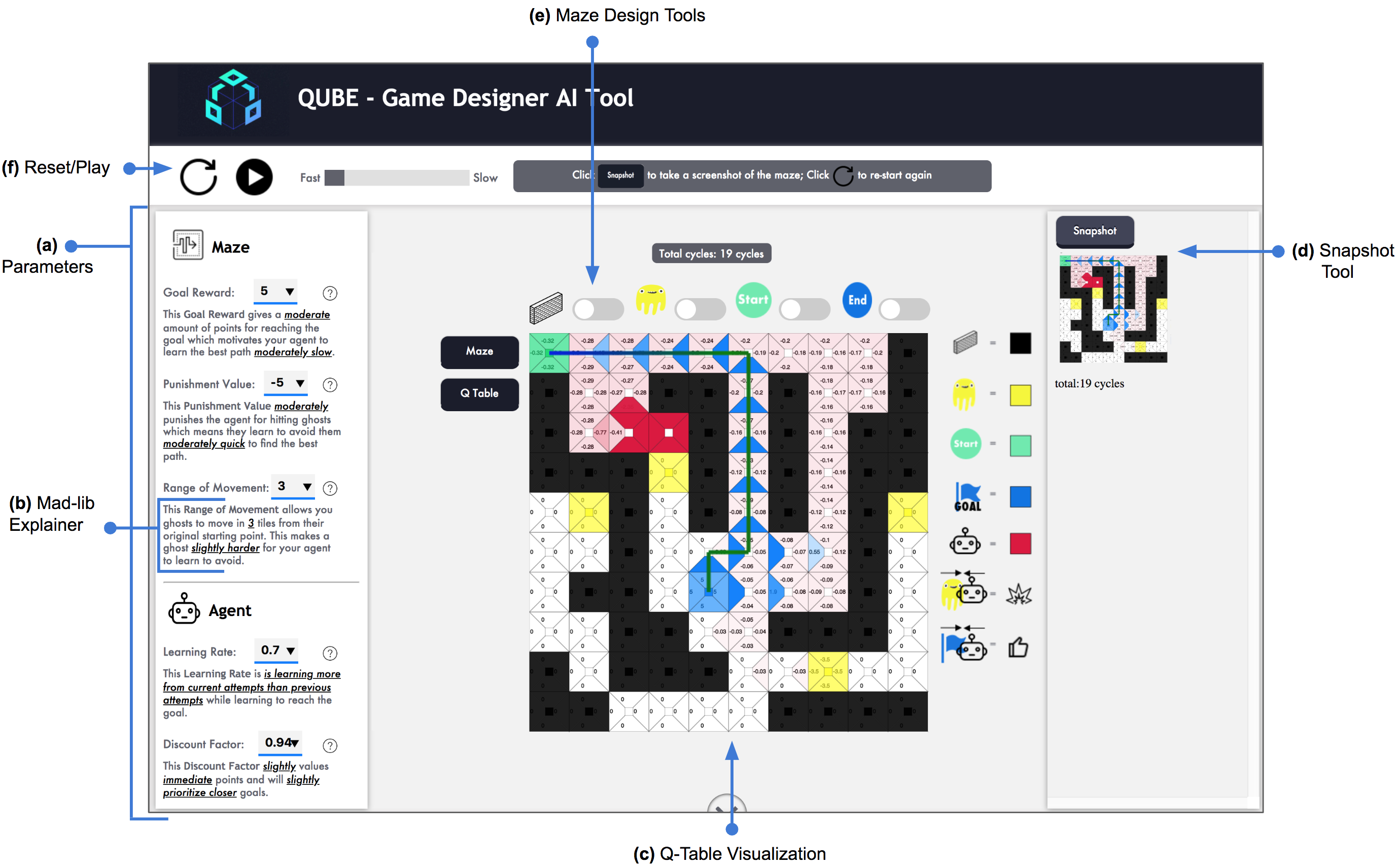}
    \caption{\textit{QUBE} with Q-Table visualization and a solved maze. (a) Q-Learning parameter users can edit. (b) An example mad-lib explainer for the range of movement. (c) The Q-Table visualizing the agent cycling through the maze. Here we see a completed/calculated path through the maze. (d) The snapshot tool with example snapshot and the cycles the agent went through to calculate its path. (e) The available tools to design the maze. (f) The controls for the simulation (reset, play, and speed). }
    \label{fig:qube_qtable_solved}
    \squeezeup
\end{figure*}

Information visualization has been shown to facilitate algorithmic learning in formal computer science classrooms~\cite{schweitzer2007interactive}. Among different approaches to educating non-experts ML concepts, simulation-based interactive visualization is widely adopted (e.g., {\em Google's} {\em TensorFlow Playground}~\cite{smilkov2017direct}). These web-based platforms primarily visualize the process of ML algorithms; the formal explanations of the principles of the processes are secondary, if they exist. These systems share three key common design elements: 1) unguided exploratory learning in a sandbox environment, 2) simulation-based interactive visualization with changeable parameters, and 3) abstract domain of algorithmic process. Our interface includes both the first and second design elements into it's own design. For the third, we abstracted the domain of the algorithmic process using a game metaphor. We hypothesized that evaluating an existing interactive visualization with designers would result in poor results since these systems are not crafted for this audience. To better tailor to our target user of digital designers, {\em QUBE}  embeds the ML process in the context of the design task of designing a game level. Game-based learning is frequently used to help students learn a wide variety of concepts~\cite{Henriksen2014WhatProcesses} such as design~\cite{Iversen2002DesignSetting}, CSS~\cite{Kim2019UnderstandingLayout}, and computational thinking~\cite{Ventura2015DevelopmentProgramming, Chao2016ExploringEnvironment,Kazimoglu2012LearningGame-play}. In programming and computational thinking games, a common game metaphor is to allow users to control a ``robot'' or ``agent'' and help this agent accomplish tasks or solve puzzles~\cite{Law2016PuzzleThinking, Chao2016ExploringEnvironment, Ontanon2017DesigningProgramming}. Of course, this application comes with new challenges. In the previously discussed game-based learning systems, the user is in direct control of the agent and its actions. If the agent is powered through ML, the user has less control over its behavior and the explanations behind it's behavior are less transparent. In this paper, we observe what benefits and detriments this agent game metaphor may bring when applied a more autonomous RL agent.

%--------------------------------------------
% METHODOLOGY
%--------------------------------------------

\section{Methodology}
{\em QUBE's} serves as an environment to evaluate how designers approach an RL concept through interactive visualizations and a game metaphor. Its design was modeled after popular interactive visualizations for ML combined with the common game-based learning metaphor of an agent solving a game ``level.'' A total of two expert panels with UX, HCI, and ML experts were conducted to evaluate \textit{QUBE} before our formal user study to ensure UX best practices were included, and RL concepts were accurately portrayed. In this section, we present \textit{QUBE'S} design, our user study design, and data analysis.

\subsection{Designing QUBE}
The design of {\em QUBE} mirrors key features of related platforms such as {\em TensorFlow Playground} and {\em GANLab} (Fig.~\ref{fig:tensorflow}). In particular, it supports 1) unguided exploratory learning in a sandbox environment, and 2) simulation-based interactive visualization with changeable parameters. In addition, being web-based makes the tool accessible from the browser directly. Below are the key design features of {\em QUBE}. More details of the system are found in~\cite{Xie2019}. We chose RL since it is popular in applications and games~\cite{Li2019ReinforcementApplications}. Q-Learning is one of the simplest RL algorithms and thus a good starting place for non-experts. Our design in \textit{QUBE} can be directly adapted to any algorithms that learns a policy (e.g., any RL algorithm, including Deep RL and any imitation learning algorithms). Since all RL learns a policy, we can query the policy/network for every single state in our game and thus create the equivalent table to visualize. 

In terms of unguided exploratory learning, {\em QUBE} is a sandbox system for the user to experiment. The system is open-ended, allowing users to explore with no predefined structure. \textit{QUBE} visualizes a Q-Learning agent learning how to optimally move from a starting spot to an end goal inside a maze. Users can interact with \textit{QUBE} in two ways: 1) changing the Q-Learning parameters manipulating the agent and 2) changing the design of the maze the agent is completing. As a simulation-based interactive visualization system, {\em QUBE} makes parameters of the Q-Learning algorithm visible and supports easy manipulation of these parameters (Fig.~\ref{fig:qube_qtable_solved}.(a)). For instance, users can change parameters such as the numeric value of the reward given to the agent upon reaching the end of the maze, the punishment value when colliding with a ghost, etc. Users can play, pause, and control the speed of the simulation of the agent with the controls seen in Fig.~\ref{fig:qube_qtable_solved}.(f). \textit{QUBE} visualizes the learning process in two different ways: 1) Q-Table visualization and 2) maze visualization. Both visualizations show the agent cycling through the maze while the Q-Learning algorithm is being executed, encountering ghosts and ``dying,'' and finding the maze's goal. In Q-Learning, the agent initially behaves erratically (likely getting lost, or colliding with ghosts), and over time, its behavior improves until converging to the optimal path. The maze visualization only shows the agent moving through the maze. To give the designer a better intuition of the behavior of the agent, the Q-Table visualization, seen in Fig.~\ref{fig:qube_qtable_solved}.(c), shows the inner workings of the algorithm. In this way, the Q-Table visualization attempts to make the Q-Learning process transparent. This transparency is accomplished by overlaying the algorithm's internal Q-Table; a numerical table that represents what the algorithm is actually learning. Each grid tile in the Q-Table contains four numbers for each possible direction in which the agent can move. The larger the number for a given direction, the higher the reward the agent believes it will get following that direction. These values show the agent's assessment on which paths to take. The numbers are also color-coded for scanability.  In this view, users can also see most likely paths highlighted by blue arrows in the Q-Table (seen in Fig.~\ref{fig:qube_qtable_solved}.(c)). 

Different from related systems, {\em QUBE} presents Q-Learning in the context of a game metaphor. It allows users to design maze levels (Fig.~\ref{fig:qube_qtable_solved}.(e)) with a visualization showing a Q-Learning powered agent calculating the optimal path through the maze (Fig.~\ref{fig:qube_qtable_solved}.(c)). Users can design a maze on a 10x10 grid by placing maze walls, adding ``ghosts'' (i.e., enemies) that can ``kill'' the agent as punishment. After pressing play, users see the agent run through the maze repeatedly, potentially colliding with ghosts and occasionally reaching the goal until it calculates the optimal path. 

During our iterative design process, we conducted two expert panels (n=3; n=6) with UX, HCI, and ML experts. Based on their feedback, we added two additional features to help designers better understand the algorithmic process. First, we incorporated a snapshot tool (Fig.~\ref{fig:qube_qtable_solved}.(d)) to help users compare results from different parameters/mazes and reflect on them. Second, we added what we call \textit{mad-lib explainers} (Fig.~\ref{fig:qube_qtable_solved}.(b)). Inspired by the phrasal template word game {\em Mad-Libs} and XAI text explanations~\cite{Ribeiro2016}, the mad-lib explainer uses a pre-defined phrase to describe the impact of a particular parameter on the agent. For instance, the mad-lib explainer for a range of movement of 0 is \textit{``This Range of Movement allows you ghosts to move in \textbf{0 tiles} from their original starting point. This makes a ghost \textbf{easier} for your agent to learn to avoid.''}  Every time the user changes a numeric parameter, its text-based explanation updates automatically. For range of movement, updating the parameter to 5 would change \textbf{``0 tiles''} to \textbf{``5 tiles''} and \textbf{``easier''} to \textbf{``difficult''}.  In the second expert panel review, we found the mad-lib explainers were successful in illustrating the impact of the parameters. 

\subsection{User Study Design}
The user study consisted of three parts: 1) pre-session survey, 2) interaction session with \textit{QUBE}, and 3) post-session survey. Each participant completed one study session with a researcher either in a lab or remote environment via video conferencing. Participants were required not to have formal ML education and to be in a field where designing with ML was applicable.

\textit{Pre-session Survey --}
In addition to demographics, we collected the participant's self-reported programming experience using Feigenspan et al.'s validated Likert item~\cite{feigenspan2012measuring} (seen in Tab.~\ref{tab:questions}). We also designed 5 additional Likert items to capture participants' self-efficacy with ML (Tab.~\ref{tab:questions}). Participants were asked to provide their definitions for ML and Q-Learning and given the option to answer that they did not know or to provide their best guess. 

\textit{Interaction Session with \textit{QUBE} --}
The participants interacted with {\em QUBE} in a lab-setting, either in-person or remotely through video conferencing. After a researcher introduced the purpose of the tool (to teach ML concepts and specifically Q-Learning), the participants were given an open-ended task to design two mazes with \textit{QUBE}. They were given no criteria for how to design the mazes and instructed to accomplish the task in any way they desired. Mazes were completed when the participant considered it done, or the participants reached 40 minutes. We used the Think-Aloud protocol during the participants' interaction with {\em QUBE} to capture their mental process, approach to design with \textit{QUBE}, and reflection on the outcome of the agent and maze after designing it.

\textit{Post-Session Survey --}
In the post-session survey, a researcher conducted a semi-structured interview with the participant following the questions in Tab.~\ref{tab:questions}. Participants then were directed to provide definitions of ML and Q-Learning. Additionally, they were asked to explain the 5 parameters of Q-Learning that are presented in \textit{QUBE}. Participants were not allowed to look up any information. Finally, participants filled out the same self-efficacy questions to capture any potential changes.

\squeezeupsmall
\subsection{Data Analysis}
For quantitative data, the Wilcoxon signed-rank test was completed to check for a significant change of the pre- and post-session self-efficacy questions. All definitions captured were scored on a 0-4 step rubric seen in Tab.~\ref{tab:rubric} by three researchers. The final score for each definition was calculated by taking the mean of the three researchers' scores (large differences in scores were discussed). Pre- and post-session ML and Q-Learning scores were compared with paired t-tests. 
For the qualitative analysis, each session was video recorded and transcribed. A thematic analysis approach was taken to analyze the recordings. Separate codebooks were developed by one researcher reviewing all transcripts and videos and two more researchers reviewing separate halves of the participants' transcripts and videos. The three researchers met and discussed their codebooks and developed one joint codebook. An inter-coder reliability (ICR) check was conducted to check for an agreement of the application of the codebook with 3 session transcripts (rounded up from 10\% of $n=21$) and their related videos. A mean ICR percentage agreement score of 92.33\% was achieved with the individual transcripts receiving a score of 91.14\%, 93.15\%, and 92.72\%. After the agreement was checked, the transcripts were split evenly amongst the three researchers, and each researcher was assigned to code each again by reviewing the transcripts, videos, and participant's definitions. Finally, the codes and their applications were reviewed jointly to develop the final themes.

% IRR
% 101 0.9114583333
% 104 0.9315770609
% 117 0.9272727273

\begin{table}[]
\centering
\setlength\tabcolsep{1pt}
\scriptsize
\begin{tabular}{|l|L{6cm}|L{1cm}|L{1cm}|}
\hline
            & \textbf{Pre-/Post- Session Self-Efficacy Survey Questions} \textit{(scale 1-5; 5 = ``Strongly Agree'')} & \textbf{Pre} & \textbf{Post} \\ \hline
\textbf{Q1} & I am confident in my understanding of machine learning.*  & 2.54 $\pm$ 1.08 &  3.52 $\pm$ 0.93             \\ \hline
\textbf{Q2} & I am confident in my ability to understand more about machine learning. & 4.0 $\pm$ 0.89 &  4.4$\pm$ 0.74\\ \hline
\textbf{Q3} & I want to improve my understanding of machine learning.   & 4.42 $\pm$ 0.67  & 4.33 $\pm$ 0.79\\ \hline
\textbf{Q4} & I want to be able to design with machine learning techniques. & 4.0 $\pm$ 1 & 4.0 $\pm$ 1.09 \\ \hline
\textbf{Q5} & I want to be able to program machine learning techniques.  & 3.76 $\pm$ 1.13 & 3.66 $\pm$ 1.01             \\ \hline

            & \multicolumn{3}{L{8cm}|}{\textbf{Pre-Session Programming Experience~\cite{feigenspan2012measuring}}}                                             \\ \hline
\textbf{} & \multicolumn{3}{L{8cm}|}{On a scale from 1 to 10, how do you estimate your programming experience? \textit{(1=``Very Inexperienced'' and 10=``Very Experienced'')}}  \\
\hline

            & \multicolumn{3}{L{8cm}|}{\textbf{Semi-Structured Interview Questions}}                                   \\ \hline
\textbf{Q1} & \multicolumn{3}{L{8cm}|}{In your own words, what is the purpose of this tool?}                \\ \hline
\textbf{Q2} & \multicolumn{3}{L{8cm}|}{What is your understanding of the Q Learning? What helped you come to that understanding?}                 \\ \hline
\textbf{Q3} & \multicolumn{3}{L{8cm}|}{Did you notice the mad-lib explanations text? What do you think of this text?}                 \\ \hline
\textbf{Q4} & \multicolumn{3}{L{8cm}|}{What features did you find the most useful when using \textit{QUBE}?}                 \\ \hline
\textbf{Q5} & \multicolumn{3}{L{8cm}|}{What features do you find the least useful or difficult to understand?}                \\ \hline
\textbf{Q6} & \multicolumn{3}{L{8cm}|}{What other improvements should be made to the tool?}                 \\ \hline
\textbf{Q7} & \multicolumn{3}{L{8cm}|}{What was useful in \textit{QUBE}'s tutorial? What was not?}            \\ \hline
\textbf{Q8} & \multicolumn{3}{L{8cm}|}{What are your thoughts on the parameters and their descriptions in the tutorial?}          \\
\hline

\end{tabular} \\
\textbf{ *$p = 0.004$ }
\caption{Self-efficacy questions, their mean values, and STD. Followed by the programming experience Likert item and the semi-structured interview.}
\label{tab:questions}
\squeezeup
\end{table}

%--------------------------------------------
% RESULTS
%--------------------------------------------
\section{Findings}
In this section, we review our quantitative results and thematic analysis findings relevant to participant behavior in a sandbox environment, interacting with a simulation, manipulating parameters, and interacting with \textit{QUBE} game metaphor.

\subsection{Participants}
A total of 23 participants were recruited from a U.S. university in a major mid-Atlantic city. Two participants' data were excluded since one reported to have formal ML training, and the other did not follow the study protocol. The mean age reported is 23.8 $\pm$ 2.93 with 14 males, 6 females, and 1 non-binary participant. Participants' academic or professional focus were recorded as game design ($n=8$), digital media design ($n=8$), animation ($n=5$), HCI research ($n=1$), and web development ($n=1$). Programming experience reported on a scale of 1-10 (10 being very experienced) had a mean score of 5.38 $\pm$ 2.37. Of the participants' whose data was analyzed, none reported having formal ML education through courses or training.

\subsection{Pre- and Post-Session Surveys}
In this section we compare pre- and post-session Likert survey scores and definition scores. 

\begin{figure}
    \centering
    \includegraphics[width=7cm]{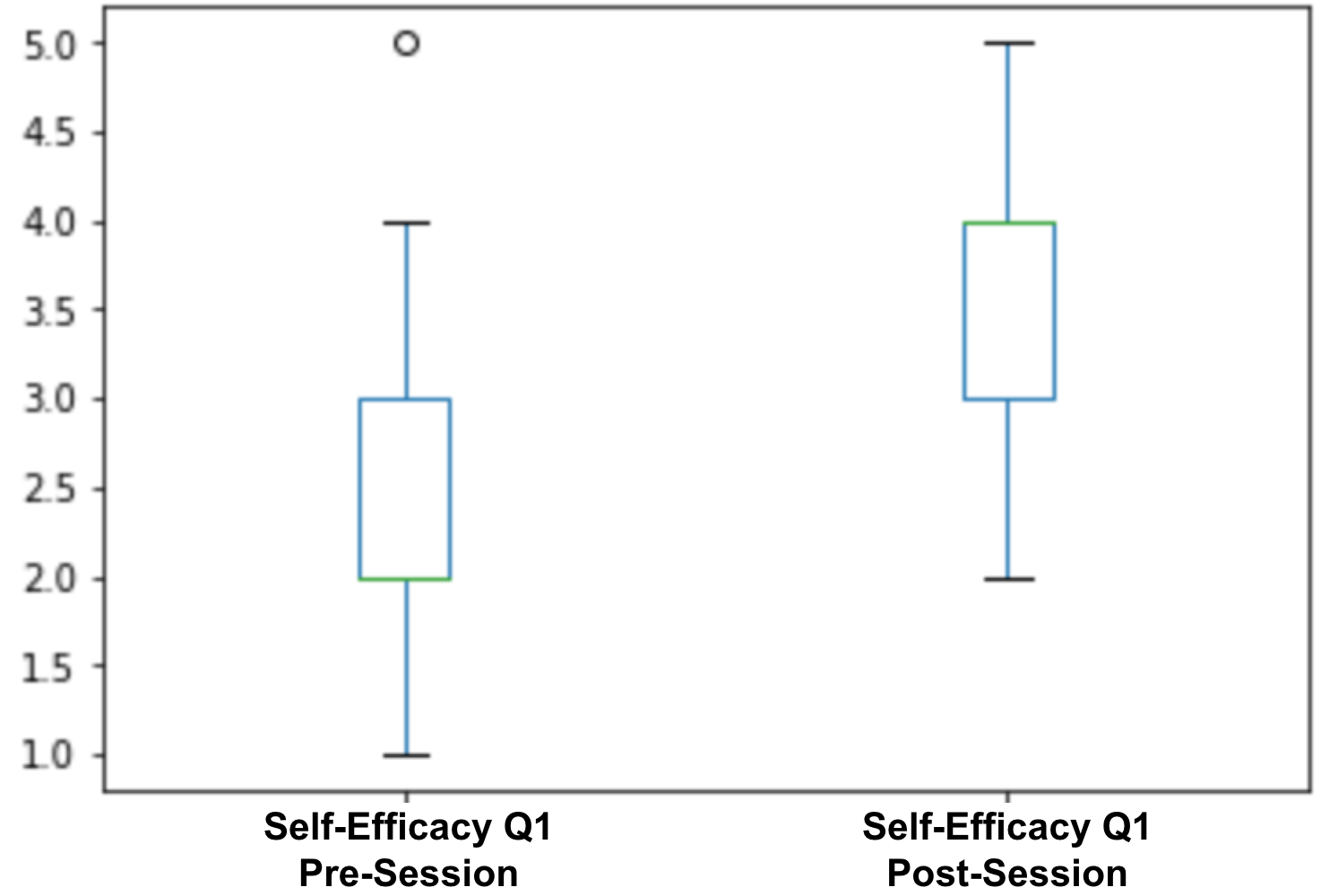}
    \caption{Boxplot of responses on a 5-point Likert scale to Q1: ``I am confident in my understanding of machine learning'' before and after the session.}
    \label{fig:q1}
    \squeezeup
\end{figure}

\subsubsection{Self-Efficacy}
Statistical significance ($Z=12.0,  p=0.004$) was found between participants' responses when asked to rate ``I am confident in my understanding of machine learning'' on a scale of 1-5 (5 being strongly agree). Distribution is represented in Fig.~\ref{fig:q1}. We recognize this increase in ML understanding could be caused by social bias since the tool was transparently meant for ML education. However, in the next section we review the change in our participants' definitions of ML. We found no statistical significance for self-efficacy Q2-5 comparing pre- and post-session responses (seen in Tab.~\ref{tab:questions}), however, the responses were generally high for each.
% +self-efficiacy
% +knowledge gain

\subsubsection{Definitions}
As discussed, participants were asked to provide definitions for ML and Q-Learning before and after the interaction session with \textit{QUBE}, which were scored by our researchers. We found that both definition scores (determined by the rubric in Tab.~\ref{tab:rubric}) significantly improved. 

\begin{figure}
    \centering
    \includegraphics[width=7cm]{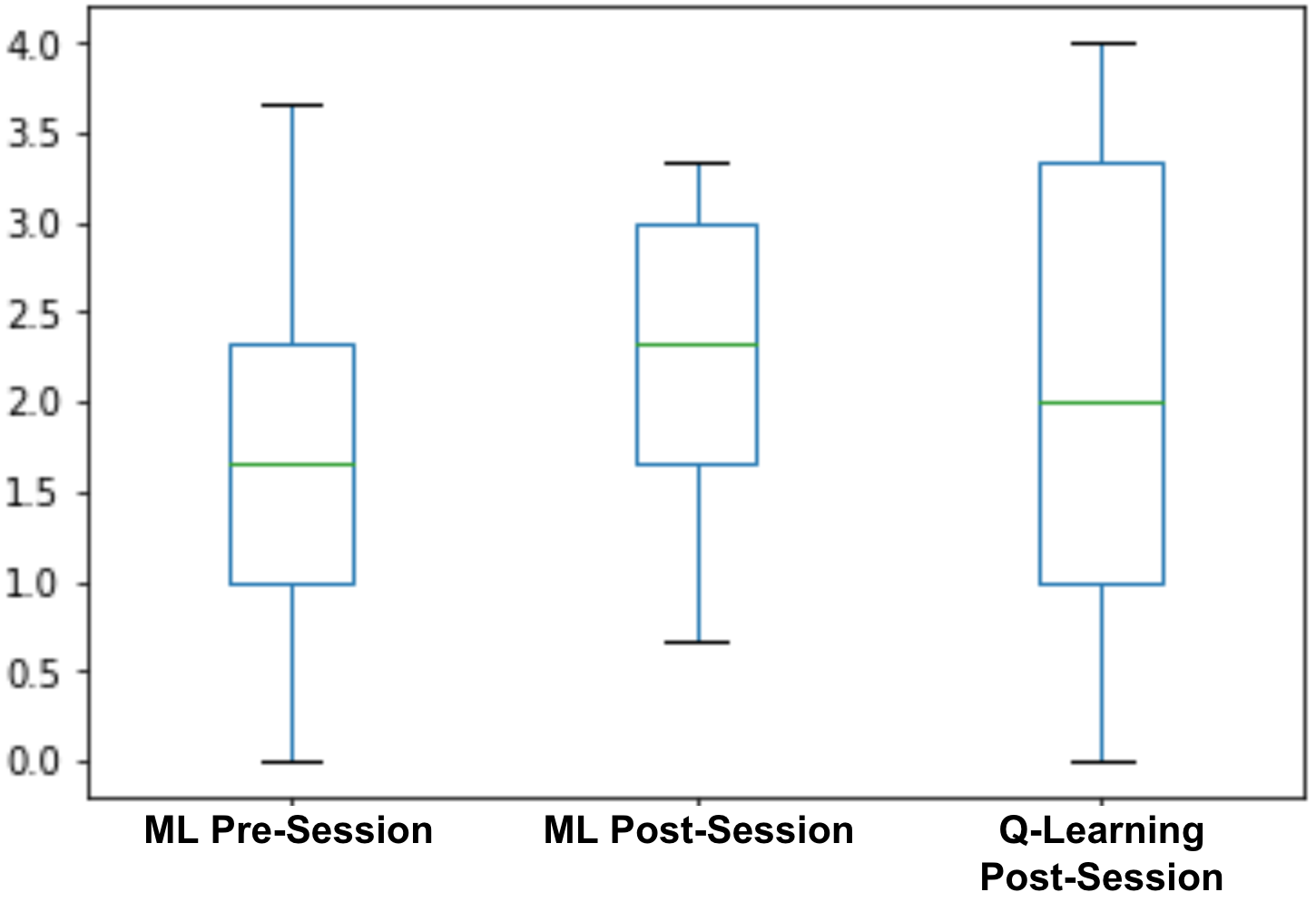}
    \caption{Boxplot of scores pre- and post-session of ML definitions and post-session Q-Learning}
    \label{fig:ml_def}
    \squeezeup
\end{figure}

\textit{ML Definition.} On a scale of 0-4 (4 being the highest score), the pre-session mean score for ML was 1.61 $\pm$ 0.97 compared to the post-session mean score of 2.26 $\pm$ 0.67 with $t(20) = -3.3$, $p = 0.003$ (seen in Fig.~\ref{fig:ml_def}). 

\begin{table}[]
\scriptsize
\setlength\tabcolsep{2pt}
\centering
\begin{tabular}{|c|L{7cm}|}
\hline
\textbf{Score} & \textbf{Rubric Score Definition}     \\ \hline
\textbf{0} & Participant (P) does not give a response or states they do not know.                                                  \\ \hline
\textbf{1} & P provides a definition that contains vaguely correct elements but is mostly inaccurate.                                         \\ \hline
\textbf{2} & P provides an almost-correct broad definition to the term but also includes some incorrect information. \\ \hline
\textbf{3} & P provides a correct broad definition of the term but does not go into details.                         \\ \hline
\textbf{4} & P provides a correct broad definition and a correct example/details. \\ \hline
\end{tabular}

\caption{Definition Scoring Rubric.}
\label{tab:rubric}
\squeezeupsmall
\end{table}

\textit{Q-Learning Definition.} For Q-Learning, post-sessions scores significantly improved ($t(20) = -7.87$, $p = 1.49\mathbf{\mathrm{e}{-7}}$) since there were no correct definitions provided pre-session. The post-session scores have a mean score of 2.08 $\pm$ 1.21. Here we see a distribution of the responses through the standard deviation showing some participants improved their understanding of Q-Learning more than others (seen in Fig.~\ref{fig:ml_def}). 
    
% \begin{figure}
%     \centering
%     \includegraphics[width=7cm]{figures/ql_def.png}
%     \caption{Boxplot of scores pre- and post-session of Q-Learning definitions}
%     \label{fig:ql_def}
% \end{figure}

\textit{Parameters Definitions.} No significance was found comparing the scores for each parameter. The definitions had the following mean scores: goal reward 2.46 $\pm$ 1.17, punishment value 2.62 $\pm$ 1.34, range of movement 2.71 $\pm$ 1.46, learning rate 2.52 $\pm$ 1.38, and discount factor 2.04 $\pm$ 1.28.

% \begin{figure}
%     \centering
%     \includegraphics[width=7cm]{figures/factors.png}
%     \caption{Boxplot of scores for the parameters}
%     \label{fig:factors}
% \end{figure}

\subsection{Thematic Analysis of Sessions and Definitions}
Due to the lack of strong quantitative findings, we conducted a thematic analysis to observe the obstacles our participant designers faced when learning the present RL concepts. Next, we review themes found when preforming a thematic analysis by reviewing each session transcript, video, semi-structured interview, and definitions. Our themes present how our participants interacted with interactive visualization design elements such as the sandbox environment, interactive simulation, and changeable parameters. Our findings also include observations on interaction with \textit{QUBE's} game metaphor.

% \textbf{\textit{RQ1: How do designers approach interactive visualization environment to form understandings of ML concept?}}

\squeezeup
\textbf{\subsubsection{\textit{Structuring Exploration in Sandbox Environment}}}
Our participants designed their own experiments when structuring their exploration of the sandbox environment. Participants were observed to gradually increase the complexity of their experimentation and design their own goals. Participants expressed desired to isolate the effect of maze design features and parameters, challenge the agent, and design the fastest agent possible. %Additionally, we observed participants heavily relied on color to interpret the interactive visualization.

\paragraph{Gradual Increase in Experimentation Complexity}\label{gradual}
% While our quantitative analysis showed that participants parameter definitions scores were not statistically different we did observe a difference in what parameters participants were more likely to initially experiment with. Additionally, w
% We observed a difference in what features of \textit{QUBE} participants experimented with first and how they selected what to experiment with next. 
Participants generally would first experiment with the least complex features. The visibility of effect determined complexity. The more visible the effect of a maze design feature (e.g., placing ghosts) or parameter (e.g., range of movement) was, the less complex it was considered by the participant. P15 commented on his behavior exhibiting this pattern: 

% \small
\textit{``[Range of movement] has the clearest change...So like the, if it's a zero, the ghost doesn't move, it's three times as well, it's a very clear, a behavioral change, that I can just tell, like, from looking at my level, how that's going to impact it.''} [P15]

% \normalsize
P15 expressed he experimented with the range of movement first because its effect was the most visible. Once participants felt they understood these less complex features and parameters, they would gradually increase the complexity of their experimentation. 
% P13 realized this pattern in their own behavior as well. 
% \textit{``It'd be interesting to see how everything plays out with the [parameters] changed up a little bit. So I understand the importance of like each number to like, make the decisions for the agent. But I guess I focused more on the design of the [maze] itself.''} [P13]
It is important to note that the gradual change occurred at different speeds for each participant. 5 participants did not alter the parameters at all once passed the tutorial and instead, similar to P13, focused on the maze design. All these participants, however, set each parameter during the interactive tutorial. P7 commented on his choice to ignore the parameters, \textit{``I didn't really feel the need to experiment with those [parameters] as much as changing the level.''} We observed participants motivations to increase the complexity of their experimentation fell into two categories: 1) participants felt they were responsible to explore more complex features or 2) participants were facing an issue with their current maze design and looked towards more complex features for an explanation. The former motivation comes from a participant feeling it was their responsibility to explore more of the interactive visualization to improve their understanding of the concepts presented. 
% \textit{``And then I'm going to try some different learning rates and discount factors. And change the rewards and stuff...I haven't mastered those as much as I should.''} [P4] 
The latter motivation was expressed when a participant's maze design was not working as expected or failing to meet their self-defined goals (e.g., an agent failing to complete a maze quick enough). 
% \textit{``The agent was dying every time. So I think the range of I can change the range of motion, right?''} [P2] 
% The participant then looked towards the more complex features to correct their agent and maze design if no solution was found editing the less complex features. 

\squeezeupsmall
\paragraph{Self-Defined Goals}\label{goals}
All participants expressed their self-defined goals when interacting with \textit{QUBE}. We categorized these goals as \textit{experimentation}, \textit{challenging the agent}, and \textit{speed}. Participants' goals could shift throughout their interaction, or they could have multiple goals at once. 17 participants expressed their goal was experimentation, 15 expressed speed, and 14 expressed challenging the agent. 

\begin{figure}
    \centering
    \includegraphics[width=9cm]{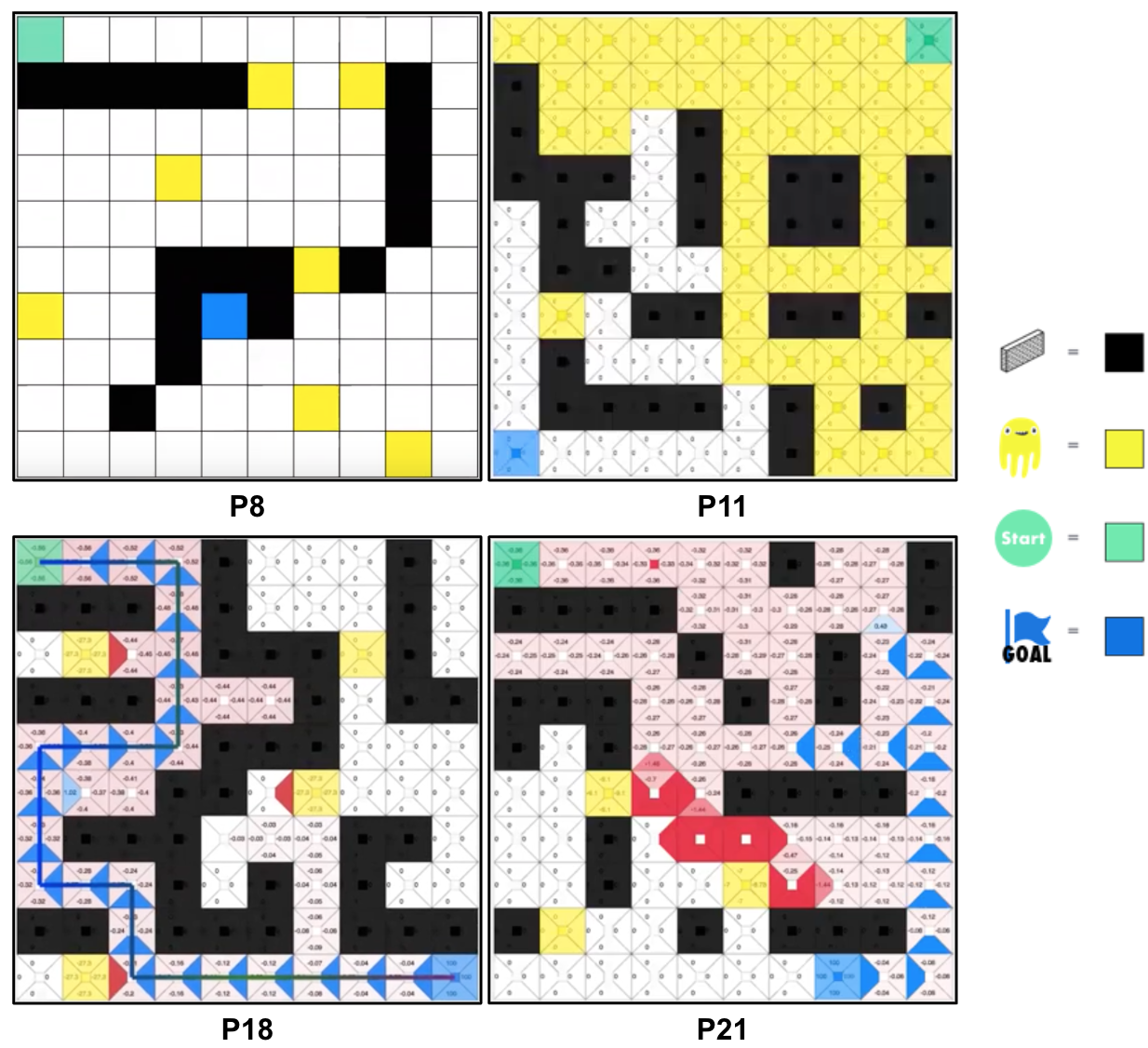}
    \caption{P8 (goal of experimentation). P11 (goal of experimentation). P21 (goal of challenge). P18 (goal of speed). }
    \label{fig:p8}
    \squeezeup
\end{figure}

\textit{Goal Experimentation.} Participants structured their own exploration by designing and executing experiments. Specifically, 11 participants explicitly reported their desire to isolate the effect of a feature or parameter.
% For example, P8 isolated the impact of the punishment value on the agent's exploration around ghosts. \textit{``I lowered the punishment value because I wanted to see if the agent would try to pass through that area.''}  
% We observed this behavior in two forms. Participants kept the parameters and changed the maze design or kept the maze design and changed the parameters to observe the effect. 
% For an example of the latter, P16 said \textit{``I'm keeping the same form [maze design] because I want to keep testing it in the same vein with different things [parameters] to see how it performs.'' } 
% 8 participants noted this process of reflection was important in their process and/or used the snapshot tool for their own reflection needs. 
% During this process, it was important for participants to be able to reflect on the state of the previous attempts to compare to the current state. The snapshot tool in \textit{QUBE} was used to achieve this. Participants most commonly would compare the path outlined by the agent or the number of cycles the agent took to come to this path. In addition to isolating specific maze design features or parameters, 
Participants desired to isolate the behavior of the agent overall. For instance, while designing their maze, P8 commented, \textit{``I was actually curious to see if the algorithm would take the path of least resistance or if it was going to go for the shorter route.''} Participants would design mazes with several paths to isolate the agent's behavior and see which one the agent deemed optimal. In Fig.~\ref{fig:p8}(P8), we can see P8's maze design with their two paths for the agent to explore. Additionally, we see in Fig.~\ref{fig:p8}(P11), the maze of P11 who expressed his desire to see how agents interacted with ghost. To do this, he designed a ``ghost wall'' to isolate the effect of ghosts on the agent. 

% P8 (goal of experimentation) wanted to observe if the agent would take the path of ``least resistance'' (right side) or the harder path (right side) to the goal. P11 (goal of experimentation) wanted to see how the agent interacted with ghosts and created a ``ghost wall'' to observe their effect. P21 (goal of challenge) edited his maze design after seeing the agent start to figure out his maze too quickly. P18 (goal of speed) selected the highest goal reward to motivate the agent to complete the maze faster. 

%   Still, 6 participants commented on their approach being the ``right'' or ``only'' way to do it. 
\textit{Goal Speed.} Participants were never informed of a ``wrong'' or ``right'' way to select their parameters. Often the ``right'' way was observed to be whichever solution led to the agent producing the optimal path the fastest. Fastness was seen to be a priority when designing the agent for 15 of our participants. This, however, is not the priority for all Q-Learning applications. Q-Learning agents can be designed to slow down and value exploration over speed. P18 commented on this observation, \textit{``...it seems like I should just choose the highest one to help my robot, my agent, learn very fast...I don't really have any reason to choose one I guess for the agent to learn slowly...''} In Fig.~\ref{fig:p8}(P18), we see the maze P18 created with a high goal reward and a ghost range of movement of 0 to increase the agent's speed through the level. Participants' goal for speed could also be motivated by benevolence instead of creating the ``optimal'' agent. Participants anthropomorphized their agent and wanted it to succeed.

\textit{Goal Challenging the Agent.} In contrast to the last goal, participants also expressed a goal of wanting the challenge the agent by manipulating the maze design and parameters to create challenging levels. P21 designed the maze seen in Fig.~\ref{fig:p8}(P21). While watching the simulation of the agent calculating the optimal path, he deemed his maze too easy for the agent since it was performing so quickly. He said he wanted to edit the maze, \textit{``Because I want it to be challenging to the computer...Let's make this a little harder for our agent.''} 
% Participants expressed similar sentiments, saying, \textit{``I wanted it to be a little challenging for the agent at least. So it couldn't just take a straight path to get to the goal.''} [P23] and  \textit{``I'm going to make it hard for the AI.''} [P11] 
Participants working towards challenging the agent expressed wanting to design a \textit{``fun or challenging''} [P5] level. The criteria for ``fun'' and ``challenging'' was based on what they thought a real person would enjoy and find challenging. 
% \textit{``I'm thinking like, how I would design it versus a player rather than a computer.''} [P21] 
% A maze was deemed challenging enough based on the speed the agent could calculate the path through it. %Participants made the maze more challenging by increasing the punishment value, increasing the range of movement of ghosts, adding more ghosts to the maze, and limiting an agents options through the maze.

%impact of unstructured learning

%Participants also isolated the behavior of the agent by giving it several paths to take through a maze. The madlib explainers and maze level wall design were two other ways participants isolated the impact of

% \textit{(b) Giving the Agent Paths to Isolate Effect.}

%

%reflection important
%difficulty understanding factor relationships

% \textit{\textbf{RQ2: What are the challenges designers encounter using an interactive visualization environment to form understandings of ML concepts?}}

\squeezeup

\textbf{\subsubsection{\textit{Experimenting with the Simulation}}}\label{s:sim}

% We observed some participants struggled to understand the ML process, the generalizability of the concepts presented, and the importance of the parameter values.

% \subsubsection{Lacking RL Process Understanding}\\

% Participants did not possess knowledge of the RL training process before this study. Most participants were able to experiment and figure out the process. Participants demonstrated this by remembering to reset after parameter and maze changes. 
12 participants struggled at one point with understanding the RL process of ``training'' (i.e., repeated maze attempts by the agent) to calculate the optimal path. Specifically, these participants were not aware of what previous attempts through the maze the agent was taking into consideration. When and why to ``reset'' the agent was not transparent. Participants would alter the maze or parameters without resetting the agent. When this happened, participants were confused about why the agent still behaved the same way as the previous maze and parameter design. For example, if a participant removed a ghost but did not, reset they would see the agent still avoided the area the ghost inhabited even after its removal. P15 struggled with this as he did not reset the agent after he edited the ghosts' range of movement to 0. 

\textit{``Even if they're [ghosts] stable, he [the agent] doesn't want to go down that right path, because I had it set to learn from his past mistakes, and he wasn't able to get past them...''} [P15]

P15's conclusion is inaccurate as he blames the learning rate for why the agent did not explore the path. Instead, the agent was still considering previous attempts where a ghost was moving into the said path and discouraging the agent from exploring it. Besides knowing when to reset, it was not transparent to participants what was happening as the agent was cycling through the level. P11 explicitly asked about this process and his role, \textit{``Do they [agent] learn from past attempts and stuff like that? Do I just leave it playing? Or do I have to like, restart it?''} 4 participants commented on how the process of the agent training was too fast and hard to follow. It was not clear to everyone the agent was making multiple attempts and ``learning''.

\squeezeup
\textbf{\subsubsection{\textit{Parameters of the Algorithm}}}

% All participants edited the parameters in the interaction tutorial \textit{or} IV. Two participants did not edit the parameters at all in the tutorial while 5 edited some of them. 5 participants did not edit the parameters at all in the IV but each edited all the parameters in the tutorial. Of these 5 participants, 2 expressed they were unsure if they could change the parameters after the tutorial. 

In general, participants were observed to be confused by the selected range and values for the parameters and how the parameters impacted each other.

\paragraph{Confusion on Parameters' Range and Increment}
The values we selected for each parameter were done so to keep the maze from being unsolvable and to offer a range to explore. Parameters in \textit{Tensorflow Playground} and \textit{GANLab} are constrained in this manner as well. Participants were often confused by the range and increments of values provided and commented on how they seem arbitrary. 16 participants commented on their confusion. 
% P12 was confused by the range of learning rate.  P14 tried to make sense of the goal reward increment: \textit{``We have 1, 3, 5, and 7, right? That's an increase of two. And then it jumps from 10, 30, to 100. Is that like, it's not like Fibonacci sequence or something?''}  

\paragraph{Confusion on Selecting Parameter Values}
Some participants struggled with identifying how to alter the parameters to achieve a desired effect. As discussed, participants had self-defined goals when interacting with the interactive visualization. Participants often did not know how to change parameters to accomplish these goals. 
% Understanding if changing a parameter value would yield a specific result could only be done through experimentation. For instance, in Section~\ref{s:sim}, we discuss P15 wanting to force the agent down a specific path but blamed the learning rate value for the agent's lack of exploration. This assumption was incorrect. 
Not understanding how the parameters impacted the agent or each other harmed participants in being able to create specific designs. Additionally, when participants were unsure of what parameter value to select they would select a middle value to ``play it safe'' as we heard from P17, \textit{``I think I'm stuck on this one. Let's just put the middle value and just go with it.''} Participants exhibiting this behavior often did not yet establish self-defined goals and had no direction for their parameter selection.

% 3 participants explicitly expressed this confusion on how parameters impacted each other and how altering one parameter's value would impact the effect of another. 

\squeezeup

% \textbf{\subsubsection{\textit{Unique to QUBE}}}

% Finally, we review our findings on features unique to \textit{QUBE}. Specifically, we review the impact of the metaphors in \textit{QUBE} (e.g., game domain) and the mad-lib explainers.

% \paragraph{Impact of Game Metaphor}
\textbf{\subsubsection{\textit{Impact of Metaphors}}}\label{s:meta}
Participants responded positively to the metaphors used in the interactive tutorial and the interactive visualization. 20 participants commented that a visualization or metaphor aided their understanding. Participants enjoyed learning through the abstracted interactive visualization and felt it was an approachable environment to explore ML concepts. \textit{``Visually explaining that through color and through numbers, which is really helping me understand how it's [the agent] navigating through the maze.''} [P16] 10 participants discussed how the Q-Table formed their understanding of Q-Learning. P13 highlighted this desire to learn through the exploration of visuals when trying to understand the Q-Table.

% \textit{ ``I wanted to play around that a little bit to see because I feel like I learned a little bit more visually. So once I see everything happening, it's easier for me to understand what's going on. At first, when I was reading through, I had a little bit of a grasp of the Q-Table. But now seeing a visually it makes more sense.''} [P13] 

Additionally, a total of 18 participants were observed to heavily rely on color to interpret the interactive visualization and its tutorial. The metaphor of the game domain also led to participants to anthropomorphize the agent and ghosts. 11 participants assigned human traits to entities in \textit{QUBE}. ``He probably wants to get to the end pretty quickly and avoid punishment.'' [P19] Participants also projected their own values on the agent. \textit{``So I'm a little disappointed. I wish he had picked the long road. And I don't know why exactly. That matters to me. But maybe, maybe I'm projecting sort of my own values.''}[P14] Since parameters were called ``rewards'' and ``punishments'', participants added emotion characteristics to their purpose as well. Participants were not instructed on how to reward or punish the agent. However, participants were observed to prefer to reward the agent over punishing. P10 stated, \textit{``Nobody wants to get punished.''} P18 summarized this phenomenon, \textit{``...because I made the goal reward kind of high. So I'll get him really excited about getting to that [goal].''} Participants believed the agent would be more ``excited'' by rewards rather than punishments and ensured reaching the goal was more rewarding than the punishment of hitting a ghost.

% \paragraph{Negative Impact of Metaphors}
The use of the game metaphor was verbally reported by 9 participants to help their understanding. The game metaphor made learning ML concepts more entertaining. However, we observed that the game metaphor example narrowed their understanding of Q-Learning. The presentation of Q-Learning in a path-finding game metaphor limited their generalization of Q-Learning's application. For example, P15's definition of Q-Learning post-session was \textit{``a grid-based path-finding algorithm that teaches a computer how to navigate a space better by passing it parameters that define what behavior is desired of the computer.''} Q-Learning is not only for grid-based path-finding, but our specific game domain shaped their definition. Out of 147 post-session definitions (21 participants x 7 definitions), we highlighted 17 definitions negatively impacted by the specific domain examples. This is a small portion, 11.56\% of post-session definitions, but is still a critical issue that could be exacerbated in other ML educational tools for non-experts.

\section{Discussion}
% - Meaning of values, abstract (low to high). Explain or reduce. 
% We believe our findings on behavior in a sandbox environment, interaction with simulation IV, and manipulation of parameters are generalizable to other IV with these characteristics targeting designers. These characteristics are also not unique to RL. Our findings on our mad-lib explainers and game domain are, however, unique to IVs seeking to also incorporate these features.

Overall, \textit{QUBE} was found to significantly improve our participants' high-level understanding of ML and Q-Learning. This aligns with previous research observing designers are capable of achieving this high-level understanding without formal education~\cite{Dove2017, Yang2018InvestigatingLearning} and through interactive visualization~\cite{Cheng2019}. However, as suspected based on previous literature~\cite{Kirschner2006WhyTeaching}, the unguided learning environment of the interactive visualization did not support their ability to design with these concepts effectively. In this section, we discuss our findings and our design recommendations for more guided learning interfaces for RL (and when generalizable, ML) concepts for designers.

\subsection{Highlight the Impact of the Combination of Parameters}
% We saw an increase in our participants' understanding, based on their definition scores, of RL concepts.
XAI techniques will play a crucial role in designing guided learning interfaces for ML concepts since they explore supporting experts and non-experts in better understanding ML decisions. We cannot give recommendations for specific XAI techniques to include in general for ML guided learning interfaces since the properties of the ML concept being presented will greatly dictate what needs to be explained (hence the large field of study). However, based on our findings we do have recommendations for characteristics to prioritize when selecting from the vast range of XAI techniques for guided learning interfaces. We narrow the generalization of our recommendations for the education of ML concepts that focus on supporting designers manipulating parameters (as is the focus of \textit{QUBE}). As reported in our findings, our participants designed their own goals to achieve with \textit{QUBE} and relied on editing the maze and changing parameters to achieve these goals. We saw our participants struggled with understanding how much of a difference changing a parameter would yield, how that change would impact other parameters, and how that change would assist in accomplishing their design goals. Because of these observations, we recommend ML guided learning interfaces include XAI techniques that illustrate the impact the combination of parameters.

Our participants were able to get a good understanding on the definition of the parameters through our mad-lib explainers but were not able to understand how to design with them. Our mad-libs were based on XAI text explanations which have been found to help non-experts understand ML reasoning~\cite{Ribeiro2016}. However, XAI text explanations typically focus on explaining the individual impact of features or parameters and do not demonstrate a parameter's influence on others~\cite{Feng2019, Ribeiro2016}. As an example, if a designer wanted to create a \textit{QUBE} agent that valued exploration, they could accomplish this by tweaking several parameters such as the range of movement of ``ghosts'' or editing the discount factor. However, our mad-lib explainers do not highlight this relationship between these two parameters. We observed our participants struggle with creating an accurate mental model of how these parameters influenced each other. we speculate that XAI techniques that focus on the impact of the combination features or parameters can help address this. Cheng et al.~\cite{Cheng2019} found their interactive visualization using a stacked bar chart showing a profiling algorithm's weights a profile's attributes helped non-experts comprehend the algorithm's decision. In this example, the more influential an attribute is on the model's decision, the larger the attribute is in the stacked bar chart. This visualization gives users an high-level overview on how the attributes combine to result in the model's decision. While our mad-libs explainers assisted in understanding the impact of one parameter they did not assist in understanding the relationship of the parameters to achieve a design goal. We believe techniques such as these could address designers' tendency to blame more complex parameters when the algorithm is not behaving in an expected manner (even when it might be a less complex feature causing the issue). Hence, we recommend including XAI techniques that focus on providing this high-level overview. However, this does not address designers in understanding what parameter to change to achieve their goals. That is why we also suggest that guided learning interfaces should guide users through design goals instead of through ML concept to concept.

% HCI guidelines dictate interfaces should support users in detecting and resolving errors~\cite{nielsen} and receive feedback on the consequences of their actions~\cite{Shneiderman2000}. However, how to do that when ML concepts are inherently opaque is the inspiration for XAI research and how do that in guided learning experiences is less explored. We suggest that guided learning experiences prioritizes the severity of changing parameter values as a way to 

% The type of explanation used will unfortunately depends on the ML concepts presented. For example, saliency maps may work well for RL but not unsupervised learning algorithms. These previous studies focus on giving a high-level understanding of RL which we see \textit{QUBE} is successful in as well. However, we believe our next priority and the goal of our study is to facilitate designers in designing with RL (and ML concept overall). While these previous studies give users accurate mental models of RL they do not help them design by manipulating the algorithm (through changeable parameters).

\subsection{Guide Learning in the Context of Design Goals}
\textit{QUBE} currently presents RL concepts and parameters through an incremental approach, presenting concepts one-by-one in the beginning tutorial. This is an approach we see in other programming guided learning interfaces as well~\cite{Kim2019UnderstandingLayout, Kazimoglu2012LearningGame-play}; as users progress, they are introduced to more complex concepts. However, by only providing definitions for our parameters, \textit{QUBE} did not support our designers in knowing what use cases these parameters were important. We recommend that concepts be presented in the context of design goals based on 1) designers struggle to generate practical design goals in \textit{QUBE} and in practice, and 2) presenting parameters in the context of realistic design goals aligns with usability best practices.

In practice, designers have shown difficulty in generating ML design goals. Designers are observed to spend a large amount of time defining their ML design goals with frequent fine-tuning for feasibility with the help of data scientists~\cite{Yang2018InvestigatingLearning}. Our participants also developed their own goals with \textit{QUBE}, however as expected, these goals did not encompass the full range of Q-Learning applications. That is why we recommend guided learning experiences to teach concepts in the context of design goals instead of incrementally presenting concepts in order of their complexity. For example, no participants had a goal of designing an agent that valued exploration of the maze over speed. These goals were also influenced by \textit{QUBE's} agent game metaphor. Participants valued creating fast agents and challenging mazes (like you would desire in a good game). Guiding learning through design goals could also balance this impact from these metaphors.

Additionally, we speculate presenting parameters in the context of what design goals they are crucial for could help designers understand what parameters to change to meet their own goals. For example, a guided learning version of \textit{QUBE} could present designers with different goals to explore, such as building a fast agent or an explorative one. The Q-Learning parameters could be presented as what parameters are the most crucial in successfully achieving these goal. Deciding what is ``crucial'' can be informed from existing research analyzing what parameters are important for different models and datasets~\cite{Probst2019Tunability:Algorithms} or through ML expert panels. Additionally, this approach aligns with existing usability guidelines, such as Nielson's 10 to ``Match between system and the real world.''~\cite{Nielson1995JakobHeuristics} By educating through  design goals, we provide designers with examples to ground their understanding of the parameters. 

\subsection{Limiting Designers' Assumptions}
We believe another key challenge in designing guided learning interfaces will be addressing designers' overestimation of the sophistication of the algorithm presented. We highlight this phenomenon since we believe future ML guided learning experiences need to prioritize limiting these assumptions through their design and educational content. We generalize this recommendation outside of RL since this behavior has been seen in other ML concepts~\cite{Dove2017}. We also see similar behavior with users of intelligent agents~\cite{Luger2016}.

With the abundance of research finding that game metaphors can support learning~\cite{Kazimoglu2012LearningGame-play, Chao2016ExploringEnvironment, Wilson2009RelationshipsProposals, Ontanon2017DesigningProgramming}, we see a risk in these metaphors further exacerbating the overestimation of sophistication of the ML concept presented. Our participants were walked through all parameters that impacted \textit{QUBE's} agent, however, some still assumed there were other ``unseen'' parameters influencing the agent's decisions. We believe this issue was exacerbated through our use of the common agent game metaphor since our participants' language included the anthromorophization of the agent. While storyline metaphors have been found to increase transfer of knowledge in game-based learning systems~\cite{Henriksen2014WhatProcesses}, we caution the use of these more sentient and personality driven metaphors when teaching ML concepts since we speculate they may increase the overestimation of the concept's sophistication. For example, presenting our Q-Learning agent as a person exploring a jungle or even a fish exploring caves may increase the assumption of increased sophistication and unseen parameters influencing the agent's decision. However, we do not recommend the absence of all metaphors since participants reported greatly enjoying this aspect of \textit{QUBE}. We recognize this challenge and believe this is an interesting area of research to explore the balance between using metaphors for teaching ML concepts while addressing overestimation. 

Furthermore, it is unclear if this overestimation would be absence even with the complete removal of metaphors. Our agent game metaphor is very minimal, and the agent is not presented with an intentional personality, and still we see the overestimation of its sophistication. This is why we also suggest explicitly addressing these assumptions early in the educational content of the system. We believe this is important for guided learning interfaces since recent interactive visualization actually \textit{may} have invisible factors influencing the model being presented. For example, \textit{Teachable Machine} is actually pre-trained on a dataset for image classification and this is purposely hidden from the users~\cite{Carney2020TeachableClassification} (this provides the benefit of users needing less samples to train their own models). Since users have a tendency to assume hidden influencing factors and since interfaces may or may not have them, we believe it is important to inform users if they do or do not have complete control early when interacting with the system.

\section{Conclusion \& Future Work}
This paper presents an overview of how designers structure their exploration of RL concepts in an interactive visualization and the challenges they face when exploring these concepts. We evaluate how designers approached design elements common in an interactive visualization; such as a sandbox environment, interaction simulation, and changeable parameters. Our user study (n=21), observed designers interacting with \textit{QUBE}, an interactive visualizer teaching Q-Learning in a maze design domain. We find while using the sandbox environment, our participants came up with their own goals. They generally approached more visible parameters first and attempted to understand them by isolating their effects. Participants understood the concepts at a high-level but struggled to experiment with them. We also observed a strong impact by the game metaphor used in {\em QUBE} on participants' understanding of RL. Based on our findings, we present design recommendations for more guided learning interfaces for RL and ML concepts.

% Designers struggled with not having enough meta-knowledge of the ML process to structure their learning, generalizing concepts learned to other domains, and in overcoming there is a ``right'' or ``best'' way to design a ML concept. Additionally, we discuss how researcher can support ML education of designers by scaffolding their exploration and aiding designers in generalizing ML concepts.

This study is limited by its focus on Q-Learning, a particular type of ML, and in a maze-level design system. It is also limited in the number of participants and their domain expertise. In this study, we did not analyze the differences in behavior and needs between designers with different domain expertise (e.g., animation versus game design). However, our findings highlight the general behavior and needs of digital designers exploring an interactive visualization for ML concepts. For our future work, we aim to redesign \textit{QUBE} to scaffold a designer's exploration based on our findings here and evaluate the impact on designers who are interested in improving their ML knowledge.

% BALANCE COLUMNS
\balance{}

% REFERENCES FORMAT
% References must be the same font size as other body text.
\bibliographystyle{SIGCHI-Reference-Format}
\bibliography{ref}

\end{document}